\documentclass
[aps,amsmath,amssymb,prl,superscriptaddress,showpacs,onecolumn,preprint]{revtex4}%
\usepackage{amsfonts}
\usepackage{amsmath}
\usepackage{amssymb}
\usepackage{graphicx}%
\begin{document}
\author{D. Phelan,$^{1}$ Despina Louca$^{\ast}$}
\affiliation{University of Virginia, Dept. of Physics, Charlottesville, VA 22904.}
\author{S. Rosenkranz,$^{2}$ S.-H. Lee,$^{1}$ Y. Qiu,$^{3,4}$ P. J. Chupas,$^{2}$ R.
Osborn,$^{2}$ H. Zheng,$^{2}$ J. F. Mitchell}
\affiliation{Materials Science Division, Argonne National Laboratory, Argonne, IL 60439.}
\author{J. R. D. Copley}
\affiliation{NIST Center for Neutron Research, Gaithersburg, MD 20899.}
\author{}
\affiliation{Dept. of Materials Science and Engineering, University of Maryland, College
Park, MD, 20742.}
\author{J.\ L. Sarrao}
\affiliation{Los Alamos National Laboratory, Los Alamos, NM 87545.}
\author{Y. Moritomo}
\affiliation{Nagoya University, Dept. of Applied Physics, Nagoya 464-8603, Japan.}
\date{\today}
\title{Nano-magnetic droplets and implications to orbital ordering in La$_{1-x}%
$Sr$_{x}$CoO$_{3}$}

\begin{abstract}
Inelastic cold neutron scattering on LaCoO$_{3}$ provided evidence for a
distinct low energy excitation at $0.6\ $meV coincident with the thermally
induced magnetic transition. \ Coexisting strong ferromagnetic (FM) and
\ weaker antiferromagnetic (AFM) correlations that are \textit{dynamic} follow
the activation to the excited state, identified as the intermediate $S=1$ spin
triplet.\ \ This is indicative of dynamical orbital ordering favoring the
observed magnetic interactions. With hole doping as in La$_{1-x}$Sr$_{x}%
$CoO$_{3}$, the FM correlations between Co spins become static and
isotropically distributed due to the formation of FM droplets. The correlation
length and condensation temperature of these droplets increase rapidly with
metallicity due to the double exchange mechanism.

\end{abstract}

\pacs{61.12.-q, 71.70.-d, 71.30.+h}
\maketitle

Transition metal oxides display rich phase diagrams due to strong correlation
effects resulting from competing crystal field, spin-spin and spin-orbit
interactions. \ The rhombohedral perovskite LaCoO$_{3}$ with $R\overline{3}c$
symmetry is a prime example where by manipulating the magnetic exchange and
electron-phonon couplings either thermally or with doping \cite{raccah,
louca1, ishikawa}, unconventional transport and magnetic characteristics
emerge \cite{tokura}. The ground state of the octahedrally coordinated
Co$^{3+}$ ion is the nonmagnetic $S=0$ that corresponds to the low spin (LS),
$t_{2g}^{6}e_{g}^{0}$~configuration. \ As the gap ($\Delta$) between the
$t_{2g\text{ }}$and $e_{g}$ orbital states becomes small in the presence of a
large trigonal distortion \cite{raccah, thornton} aided by strong Co-O
hybridization \cite{korotin}, the system can be thermally excited to a
magnetic state. \ Evidence for this is provided by a broad peak in the bulk
susceptibility, $\chi$, \cite{raccah, various, saitoh, louca2} $\sim
100\ K~$that marks the onset to a paramagnetic state. The excited state can
either be the so-called $S=1$ intermediate spin (IS) state with a $t_{2g}%
^{5}e_{g}^{1}$~configuration or the $S=2$ high spin (HS) state with a
$t_{2g}^{4}e_{g}^{2}$~configuration. The two states are fundamentally distinct
and should lead to very different interactions, considering that the $S=1$
state is accompanied by orbital degeneracy and is Jahn-Teller active while the
$S=2$ state is not. \ The low temperature dependence of $\chi$ has been fit
equally well with models that consider either spin state \cite{asai1, saitoh}
making this a controversial issue for several decades \cite{various, saitoh,
louca2, asai2}. Identifying the excited state that leads to the magnetic
transition is important as it provides insights to the nature of the Co$^{3+}$
ion interactions. \ Early thermal neutron scattering measurements showed the
existence of dynamic FM correlations \cite{asai2} but the absence of studies
on the low energy excitations has hindered the critical understanding of the
transition mechanism and is the focus of the present work.

The interactions become even more complex with the introduction of charge
carriers and the mixing of Co$^{3+}$ with Co$^{4+}$ ions \cite{louca2}. \ When
holes are doped by replacing the trivalent La$^{3+}$ ions with divalent
Sr$^{2+}$ ions in La$_{1-x}$Sr$_{x}$CoO$_{3}$, a spin glass ($x>5\%$) and a FM
- metallic (FMM) state ($x>18\%$) emerge in the phase diagram but without a
structural transition as in the isostructural colossal magnetoresistive (CMR)
manganites. \ Reports on the characteristic dynamics and spatial extent of the
magnetic correlations provided limited information. \ Using the nuclear
magnetic resonance (NMR) technique that probes local slow dynamics it was
shown that the magnetic correlations develop to an FM cluster glass with
doping \cite{itoh1}. \ A more recent NMR\ study suggested coexistence of FM
and non-FM components \cite{kuhns} proposed to be driven by tendencies for
phase separation, resulting in an inhomogeneous metallic state. \ In this
paper, we focus on the spatial distribution of the FM\ correlations and probe
both their dynamic and static characteristics. \ Our results suggest that the
intimate relation of the magnetic exchange energy to the orbital ordering
tendency and electronic configuration is key to understanding the physical
nature of the undoped LaCoO$_{3}$ as well as of the La$_{1-x}$Sr$_{x}$%
CoO$_{3}$.

Using cold neutron elastic and inelastic scattering with high instrumental
resolution, we investigated the kind of spin correlations in La$_{1-x}$%
Sr$_{x}$CoO$_{3}$ with x = 0.0, 0.10, 0.15, 0.20, that prevail in the
nonmagnetic insulating through the spin glass to the FMM state. \ In undoped
LaCoO$_{3}$, a distinct low energy excitation peaked at $\sim$0.6 meV was
observed in support of the thermal activation of a zero-field split $S=1$ spin
state. \ Concomitant with the population of this magnetic state are strong
dynamic short-range FM \textit{and} weaker AFM correlations observed for the
first time that in turn provide evidence for dynamic orbital ordering
conducive to the observed magnetic interactions. \ Upon doping, the
short-range FM correlations become static while the dynamic AFM correlations
disappear. \ Direct evidence for isotropic spread of the FM correlations
indicating the formation of FM droplets is provided. The size of these FM
droplets and their correlation length increase gradually at first in the spin
glass phase and more rapidly in the FM metallic phase. Through the double
exchange (DE) mechanism, the orbital states of Co$^{4+}$ and Co$^{3+}$ within
the droplets allow for charges to propagate throughout the crystal structure.\ 

Polycrystalline samples of LaCoO$_{3}$ (30 grams) and single crystals of
La$_{1-x}$Sr$_{x}$CoO$_{3}$ with $x$ = 0.0, 0.10, 0.15, and 0.20 (grown using
the floating zone technique and $\sim$ 6 grams each) were used for the elastic
and inelastic measurements. \ The powder samples were studied using the
cold-neutron time-of-flight Disk Chopper Spectrometer (DCS) \cite{copley} at
the NIST Center for Neutron Research (NCNR) with an incident wavelength of 5.0
$\mathring{A}$ (equivalent to an incident energy $E_{i}\ $of $3.27$ meV).
\ The single crystals were measured using the cold-neutron triple-axis
spectrometer SPINS at the NCNR, with a fixed final energy of $E_{f}=3.7$ meV.
\ The instrumental energy resolution was $\Delta$E = 0.22 meV. The single
crystals were mounted in the (h,h,l) scattering plane using the pseudocubic
notation and a room temperature lattice constant a = 3.8377 \AA . \ A cooled
BeO filter was placed in the scattered beam to eliminate higher order contaminations.

LaCoO$_{3}$ was measured using the DCS spectrometer at temperatures ranging
from 10 to 300 K. \ As this system is not magnetic at low temperatures, the 10
K data were used as the background and subtracted from data at the other
temperatures to obtain the dynamical structure factor, $S(Q,\hbar\omega)$
where \textit{Q} is the momentum transfer and $\hbar\omega$ is the energy. The
$S(Q,\hbar\omega)$ was normalized in absolute units with an accuracy of 20 \%
by comparison with the nuclear Bragg intensities \cite{lee}. The
$S(\hbar\omega)$ shown in Fig. 1 was obtained by integrating from \textit{Q} =
0.5 to 2.0 \AA $^{-1}$. Near $\hbar\omega~$= 0, the data appear noisy because
the low temperature elastic peak is subtracted from the higher temperature
data. \ No observable signal is evident in the data beyond the elastic region
at the first temperature shown, 25 K, indicating that the system is still
non-magnetic at this temperature (Fig. 1a). \ With increasing temperature, the
intensity increases at low energies and consists of two components; a broad
inelastic continuum centered at $\hbar\omega=0$ and distinct excitations at
$\pm$0.6 meV \cite{phonon}. \ The low energy inelastic scattering is commonly
seen in paramagnetic phases of magnetic systems, and it continues to grow up
to about 100\ K. \ The peaks at $\pm$0.6 meV first increase in intensity and
become well defined by 50 K (Fig. 1c) but then seem to fade away above 75 K
(Figs. 1e and 1f). \ The fading is however a consequence of the powder average
and is not seen in the single crystal data discussed below. \ The solid lines
in the figure are the fits to $S(\hbar\omega)=\frac{\chi^{^{\prime\prime}%
}(\hbar\omega)}{\pi(1-e^{-\beta\hbar\omega})}$ where the imaginary
susceptibility $\chi^{^{\prime\prime}}(\hbar\omega)~$\cite{regnault} is given
by two Lorentzian functions to account for the energy width of the two
components
\begin{equation}
\chi^{^{\prime\prime}}(\hbar\omega)=\frac{\chi_{0}\Gamma_{0}\omega}{\omega
^{2}+\Gamma_{0}^{2}}+\frac{\chi_{1}\Gamma_{1}|\omega\pm\omega_{0}|}{(\omega
\pm\omega_{0})^{2}+\Gamma_{1}^{2}}~.
\end{equation}
$\omega_{0}$ was fixed at 0.625 meV, $\Gamma_{0}$ and $\Gamma_{1}$ (with
$\Gamma_{1}$ fixed at 0.197 meV) are the relaxation rates and $\chi_{1}$ is
the amplitude of the 0.6 meV excitations. \ The first term corresponding to
the inelastic continuum intensity contains $\chi_{0}$, the static staggered
susceptibility, is compared to the bulk susceptibility, $\chi$ (Fig. 2). The
almost identical temperature dependence of $\chi_{0}$ to $\chi$ clearly
indicates the magnetic origin of the two. \ Shown in the inset is $\Gamma_{0}$
that behaves linearly with temperature, $\Gamma_{0}=0.0143k_{B}T$.

The single crystal results on LaCoO$_{3}$ as a function of $\hbar\omega$ and
$\overrightarrow{Q}$ carried out at SPINS are summarized in Fig. 3. \ The
constant-$\overrightarrow{Q}$ scans in Figs. 3a and 3b performed around the FM
$\overrightarrow{Q}=(001)$ and AFM $\overrightarrow{Q}$ = ($\frac{1}{2}%
$,$\frac{1}{2}$,$0$) wavevectors show that the 0.6 meV mode is clearly present
at temperatures higher than 75 K, in contrast to the powder data of Fig. 1.
\ Also shown are constant $\hbar\omega$ = 0.6 meV scans around
$\overrightarrow{Q}=(001)$ and $\overrightarrow{Q}$ = ($\frac{1}{2}$,$\frac
{1}{2}$,$\frac{1}{2}$) in Figs. 3c and 3d from which the magnetic scattering
is separated to three distinct components: dynamic AFM and FM correlations as
well as single ion contributions that provide the $\overrightarrow{Q}%
$-independent constant term underneath the FM and AFM peaks. While the latter
shows little or no temperature dependence, the AFM and FM correlations change
considerably with temperature. \ At 55 K, a distinct peak is present at the FM
point $\overrightarrow{Q}=(001)~$(Fig. 3c) whereas at the AFM point
$\overrightarrow{Q}$ = ($\frac{1}{2}$,$\frac{1}{2}$,$\frac{1}{2}$) only the
single ion contributions are visible. At 100 K, the FM peak gains in intensity
while a clear peak appears at the AFM point as well with about half of the
intensity of the FM peak. \ We will argue below that the 0.6 meV peak arises
from single-ion transitions within the $S=1$ triplet manifold.

Upon doping with 10 \% Sr, the 0.6 meV mode disappears as shown for data
collected at 100 K for a scan performed around the FM $\overrightarrow
{Q}=(001)~$point (Fig. 3b). Also at this temperature, FM dynamic fluctuations
are still present but suppressed while the AFM ones are absent altogether
(Figs. 3c and 3d). \ The PM fluctuations although suppressed are also present.
\ With cooling, the $x=0.10$ sample starts to exhibit static short-range
FM\ ordering below 70\ K, as shown in Figs. 4a and 4b. The elastic FM peak
becomes sharp and strong whereas the PM fluctuations weaken as the Sr
concentration increases corresponding to the enhancement of static FM
correlations. \ The elastic neutron scattering mapped in the vicinity of the
$(001)$ point in the (hhl) scattering plane shows that the spatial extent of
the FM correlations is isotropic for all values of $x$ shown. The enhancement
of ferromagnetism with increasing $x$ is also reflected in the spin freezing
temperature, $T_{c}$, defined as the inflection point of the static magnetic
scattering that increases with $x$ (Fig. 4 (b) and (d)). The width of the peak
at the FM position (Fig. 4a) determined by fitting a single Lorentzian (for
the broad magnetic contribution) and a Gaussian (for the temperature
independent weak nuclear contribution at (001)) gives a direct measure of the
correlation length, $\xi$, obtained from the inverse of the
Half-Width-Half-Maximum of the Lorentzian. \ As shown in Fig.\ 4(c), $\xi$
increases abruptly but remains finite as the system enters the metallic phase
due to the close relation between ferromagnetism and metallicity in this
system. \ This indicates that the phase transition is percolative in nature
involving FM clusters accompanied by macroscopic metallicity due to the double
exchange mechanism. \ At the same time, we observed that for each
concentration studied, $\xi$ showed no temperature dependence in the ordered
phase that contradicts recent small angle neutron scattering (SANS) data using
powder samples \cite{wu}. \ This discrepancy might be due to the powder
averaging artifact and/or the fact that SANS does not clearly discriminate
elastic from inelastic contributions. \ 

Our neutron scattering results provide important information on key issues of
the physics of La$_{1-x}$Sr$_{x}$CoO$_{3}$. \ In pure LaCoO$_{3}$,
correlations between the thermally excited magnetic Co$^{3+}$ ions have both
FM and AFM characteristics. This can be explained by the Co ions in the
intermediate $t_{2g}^{5}e_{g}^{1}$, $S=1$ state and with an orbital ordering
that favors this state. The IS state is Jahn-Teller active as in the
isostructural LaMnO$_{3}$ with $S=2$, $t_{2g}^{3}e_{g}^{1}$. \ In the latter,
the degeneracy is broken through \textit{static} Jahn-Teller orbital ordering
leading to A-type AFM ordering consisting of FM interactions in the $ab$-plane
and AFM coupling between the planes coupled with a change in symmetry to
orthorhombic \cite{fujimori}. \ The coexistence of \textit{dynamic} FM and AFM
correlations in LaCoO$_{3}$ suggests that orbital ordering analogous to the
A-type ordering most likely occurs. \ We emphasize that the magnetic
correlations are dynamic and short range with $\xi\approx$3.6 \AA \ that is
close to the nearest neighbor Co-Co distance of \symbol{126}3.8 \AA . \ This
implies that the orbital correlations are also dynamic and because of their
short-range nature, they occur randomly in every direction. \ As a result,
this generates dynamic FM and AFM correlations in every direction instead of
the static and long-range A-type ordering of LaMnO$_{3}$. This can explain why
magnetic scattering is observed at every integer and half-integer points. At
the same time, the HS $S=2$ state of the Co$^{3+}$ ion is unlikely because
only AFM correlations are expected between the \textit{S} = 2 states
\cite{sawatski}. The 0.6 meV mode is most likely due to transitions within the
manifold of the thermally induced magnetic state \cite{Podl}, split in zero
field due to the trigonal lattice distortion \cite{abragam}.\ This is in good
agreement with recent electron spin resonance (ESR) studies that extracted
from the field splitting of the resonance lines that LaCoO$_{3}$ has a spin
triplet split by a uniaxial single ion anisotropy, D $\approx$ 0.6 meV
\cite{noguchi}. \ 

With hole doping, short-range static FM correlations emerge at low
temperatures. The FM correlations rapidly enhance as the system enters the
metallic phase indicating that spins couple with charge via the DE mechanism
\cite{zener}. Upon doping, magnetic Co$^{4+}$ ($t_{2g}^{5}$, S=1/2) ions form
and induce the IS spin states of neighboring Co$^{3+}$ ($t_{2g}^{5}e_{g}^{1}$)
ions \cite{louca1}, forming FM droplets. Within a droplet, charge hopping
requires the $e_{g}$ level for Co$^{4+}$ to be empty and for Co$^{3+}$ to be
in the IS state and the two ions to be coupled ferromagnetically. \ Outside
the droplet, spins couple dynamically and paramagnetically. \ When the droplet
concentration gets large enough that connectivity is established between them,
charge can then propagate over the lattice transforming the system to a
metallic ferromagnet. \ In conclusion, our results provide indirect evidence
in support of the thermally excited $S=1$ IS state and dynamic orbital
ordering in pure LaCoO$_{3}$. \ The addition of charges enhances the formation
of nanoscale ferromagnetic droplets due to local DE interactions between the
Co$^{4+}$ and the IS spin state Co$^{3+}$ ions.

Work at the University of Virginia is supported by the U. S. Department of
Energy, under contract DE-FG02-01ER45927 and the U.S. DOC through
NIST-70NANB5H1152, at the Los Alamos National Laboratory under contract
W-7405-Eng-36 and at the Argonne National Laboratory under contract
W-31-109-ENG-38. \ The use of the Neutron Scattering facilities at NIST was
supported in part through NSF grants DMR-9986442 and DMR-0086210.

$^{\ast}$Corresponding author.

Electronic address: louca@virginia.edu

Figure captions:

Figure 1: The $\hbar\omega$-dependence of the neutron scattering cross
section, S($\hbar\omega$), obtained from the DCS measurements on a powder
sample of LaCoO$_{3}$ at various temperatures: (a) 25 K, (b) 50\ K, (c) 75 K
and (d) 100K. \ 

Figure 2: The temperature dependence of the static staggered susceptibility,
$\chi_{0}$, as determined from eqn. 1 (open squares) and the bulk magnetic
susceptibility, $\chi$, (filled circles). \ The upturn below 35 K in $\chi$
might be from a surface ferromagnetic component as described in Ref.
\cite{goodenough}. \ Beyond 75\ K, the two peaks are not resolved and only one
term is used for fitting $\chi^{^{\prime\prime}}(\hbar\omega)$. \ The
temperature dependence of $\Gamma_{0}$ is shown in the inset.

Figure 3: Inelastic neutron scattering data obtained from the single crystals
of LaCoO$_{3}$ and La$_{0.9}$Sr$_{0.1}$CoO$_{3}$ at SPINS.
\ Constant-$\overrightarrow{Q}$ scans at (a) AFM $Q=(\frac{1}{2},\frac{1}%
{2},0)$ and (b) FM $Q=(001)$ wavevectors. \ Constant $\hbar\omega=0.6$ meV
scans centered at (c) AFM $Q=(\frac{1}{2},\frac{1}{2},\frac{1}{2})$ and (d) FM
$Q=(001)$ points.

Figure 4: (a) Elastic scan along (0,0,L) performed on single crystals of
La$_{1-x}$Sr$_{x}$CoO$_{3}$ with x = 0.1, 0.15, and 0.2 at 8 K. The
temperature dependence of (b) the elastic neutron scattering intensity, (c) of
the ferromagnetic correlation length, $\xi$, and (d) of the spin freezing
temperature, T$_{c}$.


\begin{thebibliography}{99}
\bibitem {phonon}The 0.6 meV mode is of magnetic origin as the intensity does
not follow the Q$^{2}$ dependence of phonon modes and a field effect has been
reported in Ref. \cite{Podl}.
\end{thebibliography}
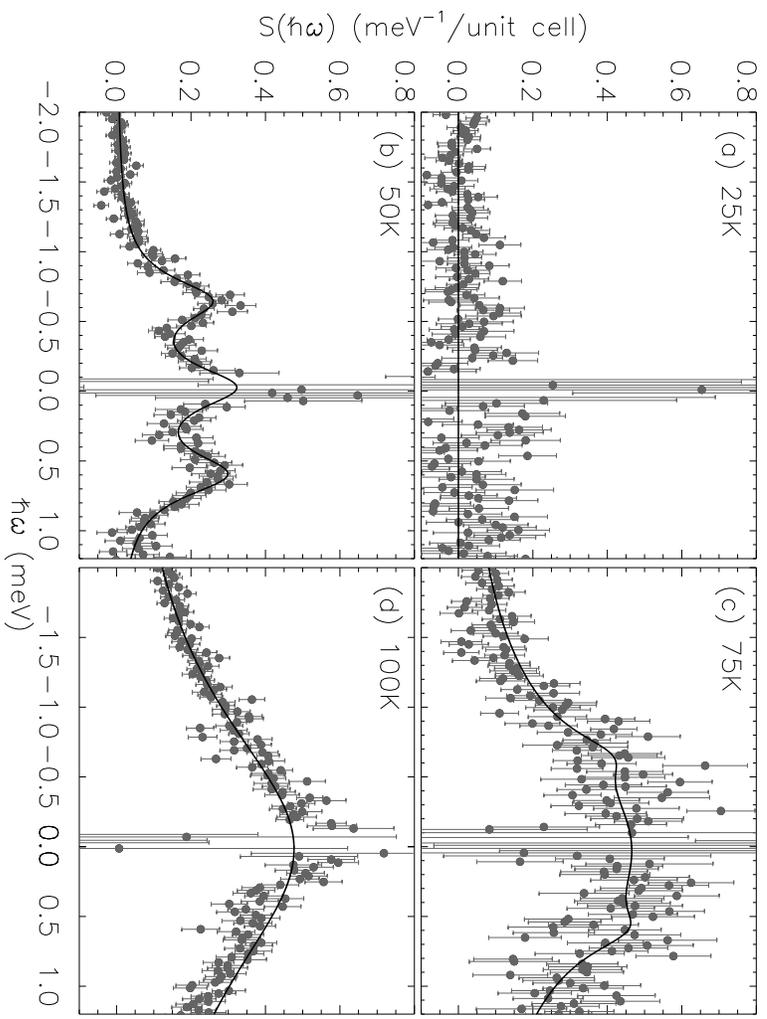
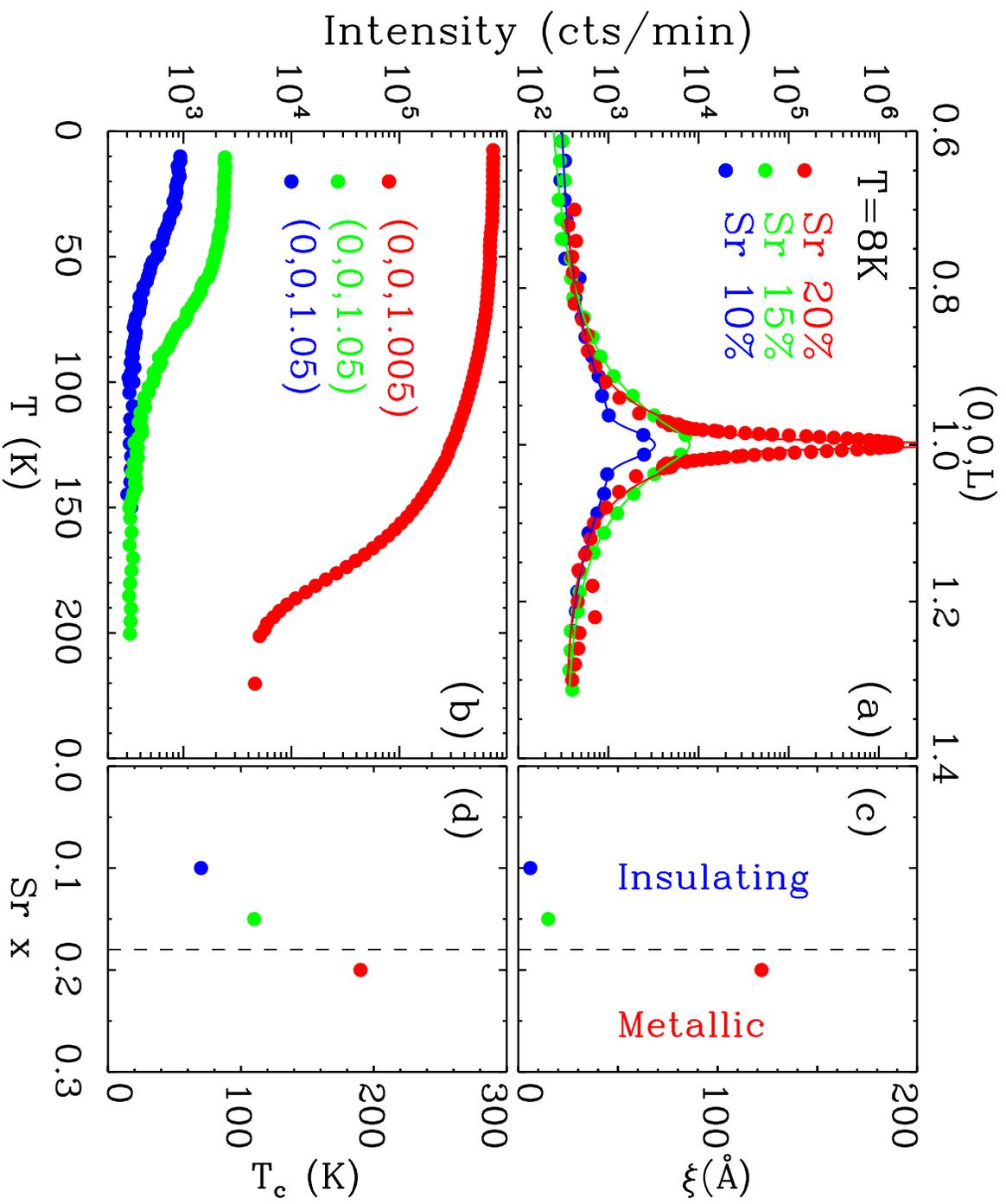
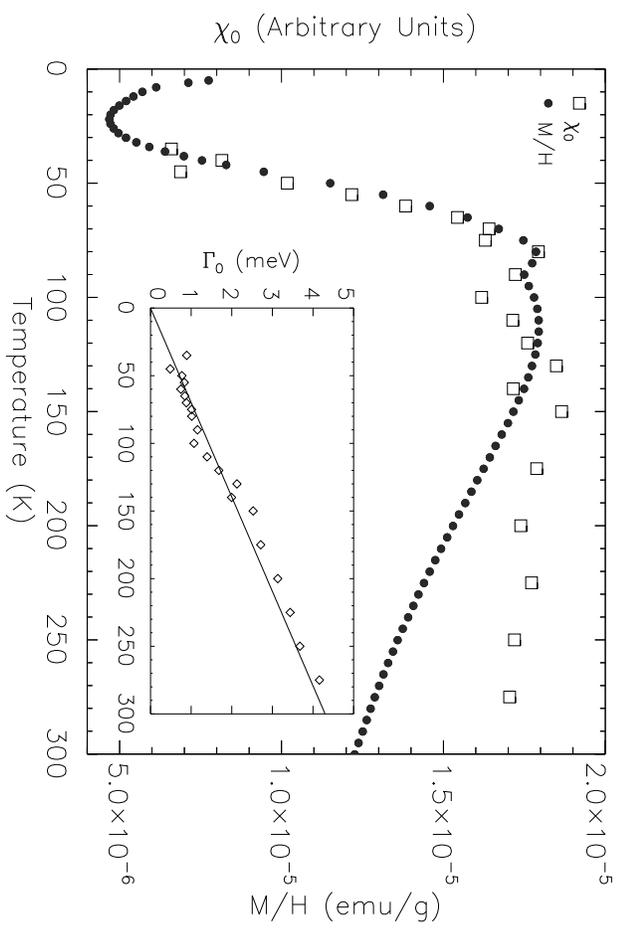
\end{document}